\documentclass[allclo]{FBSart}
\usepackage{amsfonts}
\usepackage{amssymb}
\usepackage[dvips]{graphicx}

\newcommand{\gl}{\lambda}

\newcommand{\gL}{\Lambda}

\title{The stability of the relativistic three-body system and in-medium equations}
\author{S.~Mattiello\thanks{\textit{E-mail address:} 
 stefano.mattiello@physik.uni-rostock.de}}
\institute{Fachbereich Physik,
 University of Rostock, 18051 Rostock, Germany}

\runningauthor{S.~Mattiello}
\runningtitle{}
\begin{document}

\maketitle
\begin{abstract}
We present a relativistic three-body equation to study the stability of the isolated three-body system and the correlations in a medium of finite temperatures and densities. Relativity is implemented utilizing the light front form. Using a zero-range force we find the relativistic analog of the Thomas collapse and investigate the possibility that the nucleon exists as a Borromean system.
Within a systematic Dyson equation approach we calculate the three-body Mott transition and the critical temperature of the color-superconducting phase.

\end{abstract}

\section{Introduction}

Applications of concepts of effective field theory to the quantum chromo\-dy\-na\-mics (QCD), in which the hadrons are viewed as bound system of three relativistic constituent quarks, make it useful to investigate the properties of relativistic equations of three particles interacting through effective attractive forces. In this scenario, zero range interactions provide a simple,but important limiting case
for short range forces~\cite{Fedorov:2001wj,Fedorov:2001vw,Frederico:1992uw,Carbonell:2002qs}.

The non-relativistic three-body system based on contact forces shows the Thomas collapse~\cite{Thomas1935}; this effect occurs, since the binding energy of the system is
unbounded from below. However, a relativistic treatment of the problem is necessary, if the binding energy becomes larger and eventually exceeds the size of the constituents.
The relativistic three-particle problem with a contact
interaction has been investigated using relativistic light
front equations~\cite{Frederico:1992uw,deAraujo:1995mh}. The regularization procedure that has been used avoids the Thomas collapse. Recently it
has been revisited by Carbonell and Karmanov in a covariant light
front approach without introducing an explicit regularization scheme in the
three-body equation~\cite{Carbonell:2002qs}.

 Understanding the relativistic isolated few-body bound problem is the first step to study consistently the three-quark clusters at finite temperatures and densities. 
Results of lattice calculations and model simulations suggest a rich structure of the QCD phase diagram~\cite{Karsch:2000vy,Alford:2001dt} including the transition from quarks to nucleons
as relevant degrees of freedom (Mott transition, i.e. dissociation~\cite{Beyer:2001bc,Mattiello:2001vq}), the chiral restoration and the color superconductivity analogous to Cooper pairing.
In this context three-quark correlations should play an important role in the vicinity of the phase transition and are investigated within a Green functions formalism using the Dyson equation approach~\cite{duk98}.
%

In this paper, we consider first the isolated case and then medium effects.
We study the dependence on an
invariant cut-off $\Lambda$ and show the relativistic analog of the Thomas collapse. After that we study the equations in medium and calculate for different values of the cut-off $\Lambda$ the dissociation of the three-quark bound states as well as the critical temperature $T_c$ of the color-superconducting phase.

\section{Isolated case}
For the time being we investigate bose-type relativistic equations.
Using a zero-range interaction the solution for the two-body propagator $t(M_2)$ is given by~\cite{Frederico:1992uw}
\begin{equation}
t(M_2)=\left(i\gl^{-1} - B(M_2)\right)^{-1},
\label{eqn:tau}
\end{equation}
where the expression for $B(M_2)$ corresponds to a loop diagram. In the rest
system of the two-body system $P^\mu=(M_2,0,0,0)$ it is given by
\begin{equation}
B(M_2)=-\frac{i}{2(2\pi)^3} \int \frac{dx d^2k_\perp}{x(1-x)}
\frac{1}{M_2^2-M_{20}^2},
\label{eqn:B}
\end{equation}
where $M_{20}^2=(\vec k_\perp^{~2}+m^2)/x(1-x)$ and $x=k^+/P^+_2$.
The logarithmic divergence appearing in
the integral can be absorbed in a redefinition of $\gl$. To do so one assumes
that the two particle propagator $t(M_2)$ has a pole for $M_2=M_{2B}$, i.e.
\begin{equation}
i\gl^{-1}=B(M_{2B}).
\label{eqn:Bound}
\end{equation}
The subtraction imposed by this condition in the denominator
of eq.~(\ref{eqn:tau}) makes $t(M_2)$ finite~\cite{Frederico:1992uw,Carbonell:2002qs,Beyer:2001bc,Mattiello:2001vq}. However, in
order to investigate the three-body bound state equation even if no two-body bound state exists, we use in the integral (\ref{eqn:B}) an invariant cut-off $\gL$~\cite{stab:2003}. The requirement is that the mass of the virtual two-body subsystem is smaller than the cut-off, i.e. $M_{20}^2<\gL^2$ and hence $t\rightarrow t_\gL$.
With this cut-off the integral is given by
\begin{equation}
B_\Lambda(M_2)=-\frac{2\pi i}{2(2\pi)^3} 
\int\limits_{x_{\rm min}}^{x_{\rm max}}
 \frac{dx }{x(1-x)} \int\limits_0^{k_{\rm max}}k_\perp dk_\perp
\frac{1}{M_2^2-M_{20}^2},
\label{eqn:Breg}
\end{equation}
where 
\begin{eqnarray}
x_{\rm min,max}&=&\frac{1}{2}\left(1\mp\sqrt{1-4m^2/\gL^2}\right)\\k^2_{\rm max}&=&\gL^2x(1-x)-m^2.
\end{eqnarray}

The solution of the two-body propagator is the input for the three-body equation. For the mass
of the virtual three-particle state, which is the sum of the on-shell minus-components of the three
particles, 
we introduce a similar regularization  
that leads to parametric dependence of the vertex function $\Gamma_\gL$ on the cut-off $\gL$,
\begin{eqnarray}
\Gamma_\Lambda(y,\vec q_\perp) &= &\frac{i}{(2\pi)^3}\ t_\Lambda(M_2)
\int_{0}^{1-y} \frac{dx}{x(1-y-x)}\nonumber\\
&&\int d^2k_\perp
\frac{\theta(M^2_{30}-\Lambda^2)}
{M^2_3 -M_{03}^2}\;\Gamma_\Lambda(x,\vec k_\perp),
\label{eqn:fad}
\end{eqnarray}

\begin{figure}[b]
\begin{center}
\includegraphics[width=.8\textwidth]{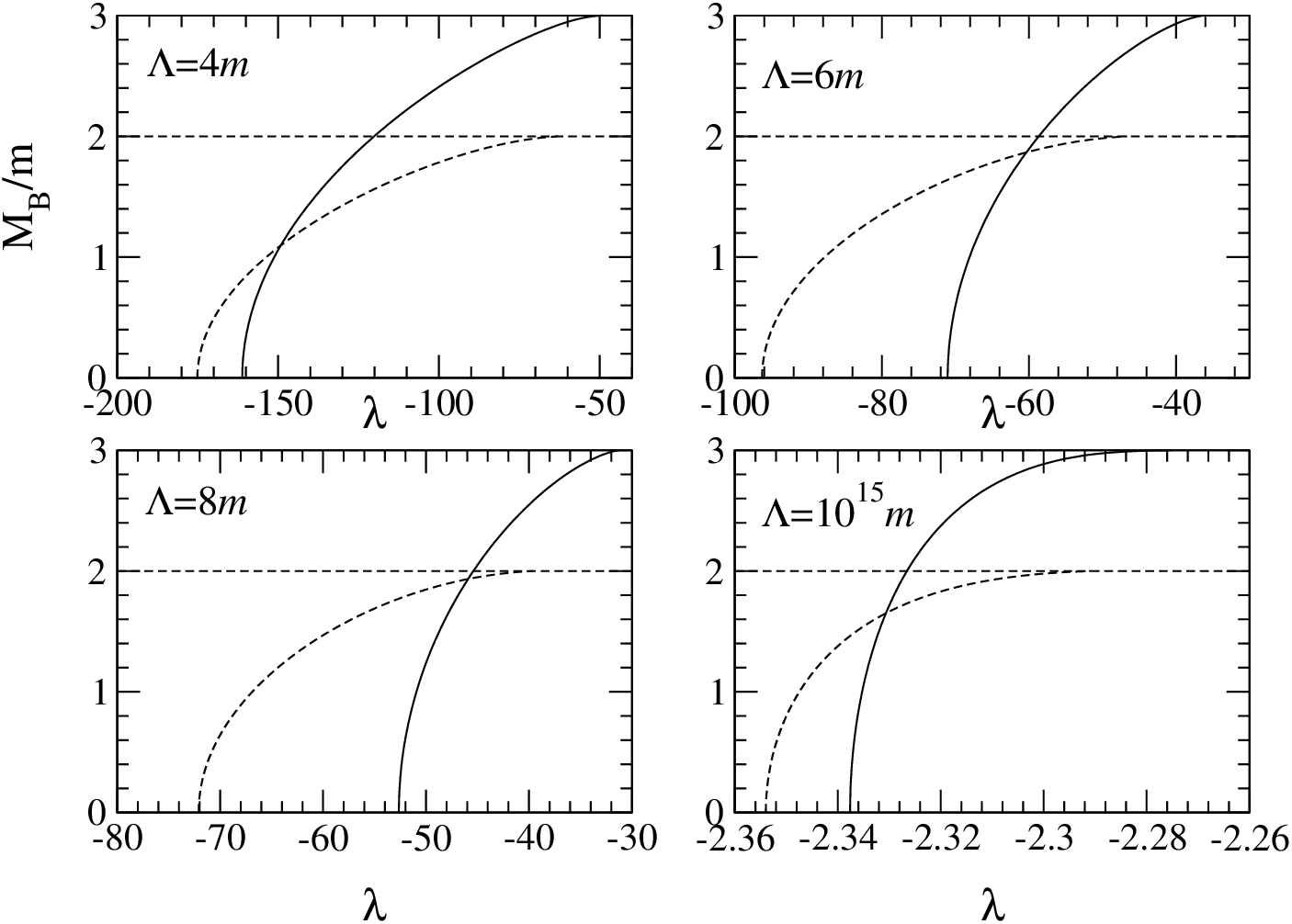}
\caption{Solution of the two- (dashed) and three-body (solid) bound state
  equations as a function of the strength $\lambda$ for
  different cut-off parameters $\Lambda=4m,6m,8m$ and
  $\Lambda=10^{15}m$ Figs. a,b,c, and d respectively.  Horizontal
  dashed lines show the two-body break-up.\label{fig:m23L}}
\end{center}
\end{figure}
\begin{figure}\begin{minipage}[t]{0.48\textwidth}
\includegraphics[width=\textwidth]{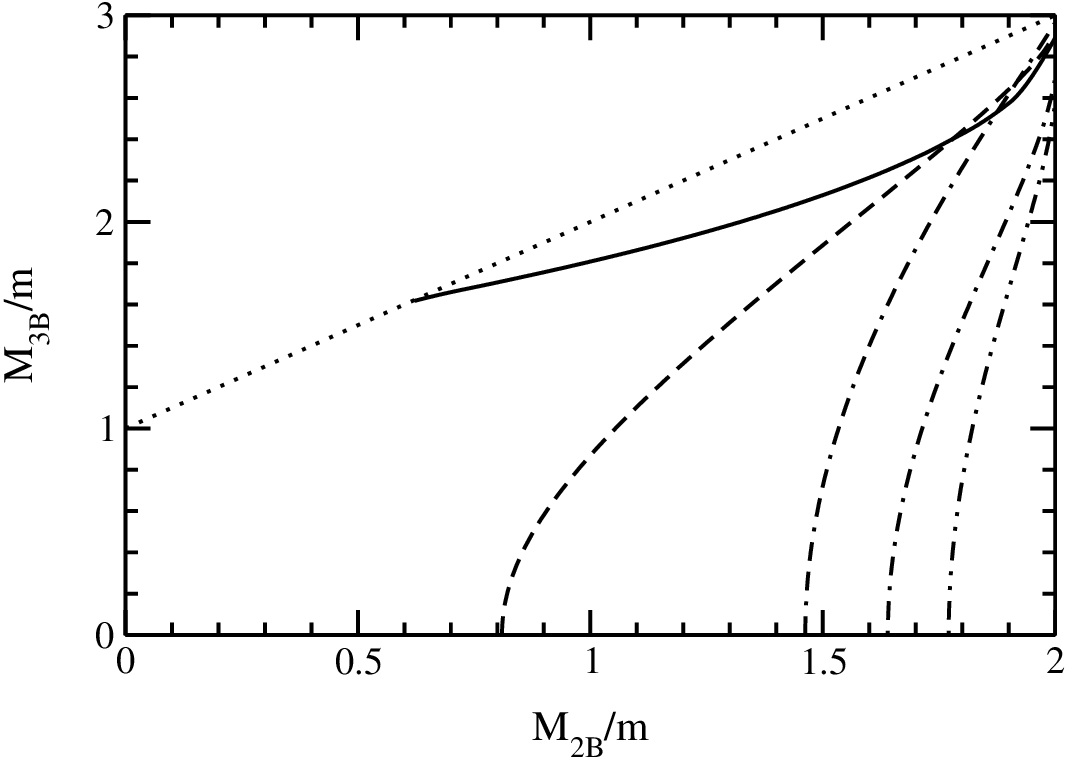}
\caption{Three-body bound state as a function of $M_{2B}$ for different
  regularization schemes. Scheme B solid line. Others use invariant
  cut-off with different $\Lambda$: $\Lambda=4m$ (dash),
  $\Lambda=6m$ (dash-dot) $\Lambda=8m$ (dash-dot-dot)
  $\Lambda\rightarrow\infty$ (dash-dash-dot).\label{fig:M23}}
\end{minipage}
\hfill
\begin{minipage}[t]{0.48\textwidth}
\includegraphics[width=\textwidth]{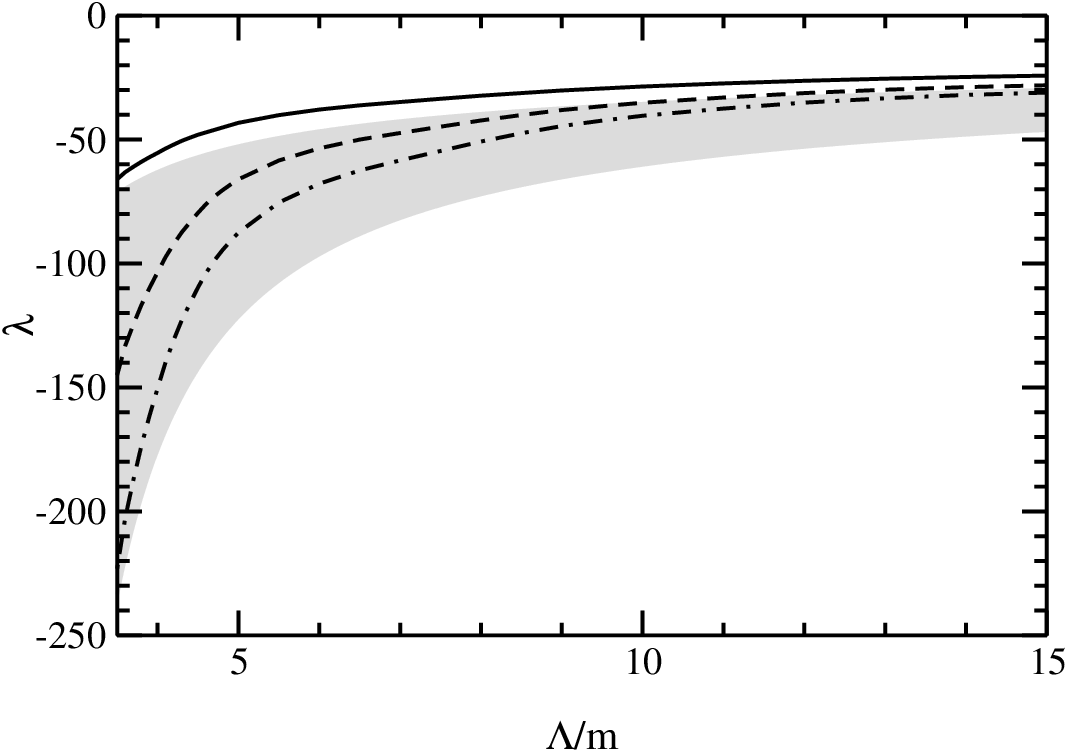}
\caption{$\lambda(\Lambda)$ from a fit of $M_{3B}$ to the proton mass, with
  $m=315$ MeV (solid line), $m=400$ MeV (dashed line) $m=900$
  MeV (dash-dot). Grey area bounded by $M_2=2m$ (upper), $M_2=0$
  (lower).\label{fig:L}}
\end{minipage}
\end{figure}
 We investigate the  two- and the three-body bound
state as a function of the strength $\lambda$ for different values of $\gL$.
To solve (\ref{eqn:fad}) we
use the cut-off parameters $\Lambda=4m,6m,8m,$ where $m$ is the
constituent mass, and $\Lambda/m=10^{15}\rightarrow\infty$.
In Fig.~\ref{fig:m23L} we show the two-body mass $M_{2B}$ and the three-body mass $M_{3B}$ in units of the particle mass. The values of the strength $\gl,$ where bound
states exist, become smaller as the cut-off $\gL$ increases.
For certain values of these parameters the three-body bound
state collapses, i.e $M_{3B}\rightarrow 0$, for a finite two-body mass. This corresponds to the nonrelativistic Thomas collapse~\cite{Thomas1935}.
However, there is a region of parameters where both $M_{2B}$ and
$M_{3B}$ exist. In this case it is possible to plot
$M_{3B}$ vs. $M_{2B}$, as shown in Fig.~\ref{fig:M23}.
The correlations between two- and three-body bound states are presented for the different cut-offs parameter chosen and for the regularization scheme, which we call B,  used in ~\cite{Frederico:1992uw,Beyer:2001bc,Mattiello:2001vq}.
This avoids the Thomas collapse, because of the additional restriction $M_2^2\ge 0$ for the intermediate state.

For weak bound two-body states the dependence of the resulting function $M_{3B}(M_{2B})$ is smooth and reproduces qualitatively in this range the results of the regularization B. However, we note that for the large invariant cut-off this function coincides with the result of Ref.~\cite{Carbonell:2002qs}.

The introduction of a cut-off allows us to solve the three-body equation without assuming a two-body bound state, as e.g. utilized by~\cite{Frederico:1992uw,Carbonell:2002qs}, but extending our regularization procedure.
Note that for a fixed $M_3$ the function $M_2(\Lambda)$ is not monotonic. This is due to the necessary readjustment of the strength $\lambda$ that eventually acts differently in the 2 and the 3-body system in combination with the cut-off. The function $\lambda(\Lambda)$ however for a given mass $M_3$ is a monotonic function as is should be to keep $M_3$ constant, see Fig.~\ref{fig:L}.

Although the model neglects important aspects of the baryon dynamics, i.e. the spin, we introduce the proton mass ($m_p=938$ MeV) as a scale of the calculation.
On the light front the treatment of the spin is technically difficult and for the time being we average over the spin projections.
This procedure can be justified in matter, because it means that the spins are washed out in the medium.
The introduction of the scale allows us to obtain a relation between the parameters $(m,\lambda,\Lambda)$.
We choose a different values of the quark mass:$m\simeq m_\rho/2\simeq
400$ MeV~\cite{deAraujo:1995mh} and $m\simeq m_p/3\simeq 315$ MeV and a
rather large one of $m=900$ MeV. The fit to the proton mass determines the functions $\lambda(\Lambda)$ that are shown in
Fig.~\ref{fig:L}. The grey area delimits the range of the parameters for which the two-quark bound states exist. The lines outside this area indicate where the proton can be found without bound two-body subsystems, i.e. the proton is described as a Borromean system. That occurs for $m=315$ MeV for all values of $\Lambda$ and for $m=400$ MeV for values of $\Lambda$ above $\sim 9.6m$.
With the choice of a certain value for the quark mass the model is parameterized by the input $\Lambda$ only. 

In the following we assume $m=400$ MeV and use it for the in-medium calculations.

\section{The problem in medium}
\begin{figure}\begin{minipage}[t]{0.48\textwidth}
\includegraphics[width=\textwidth]{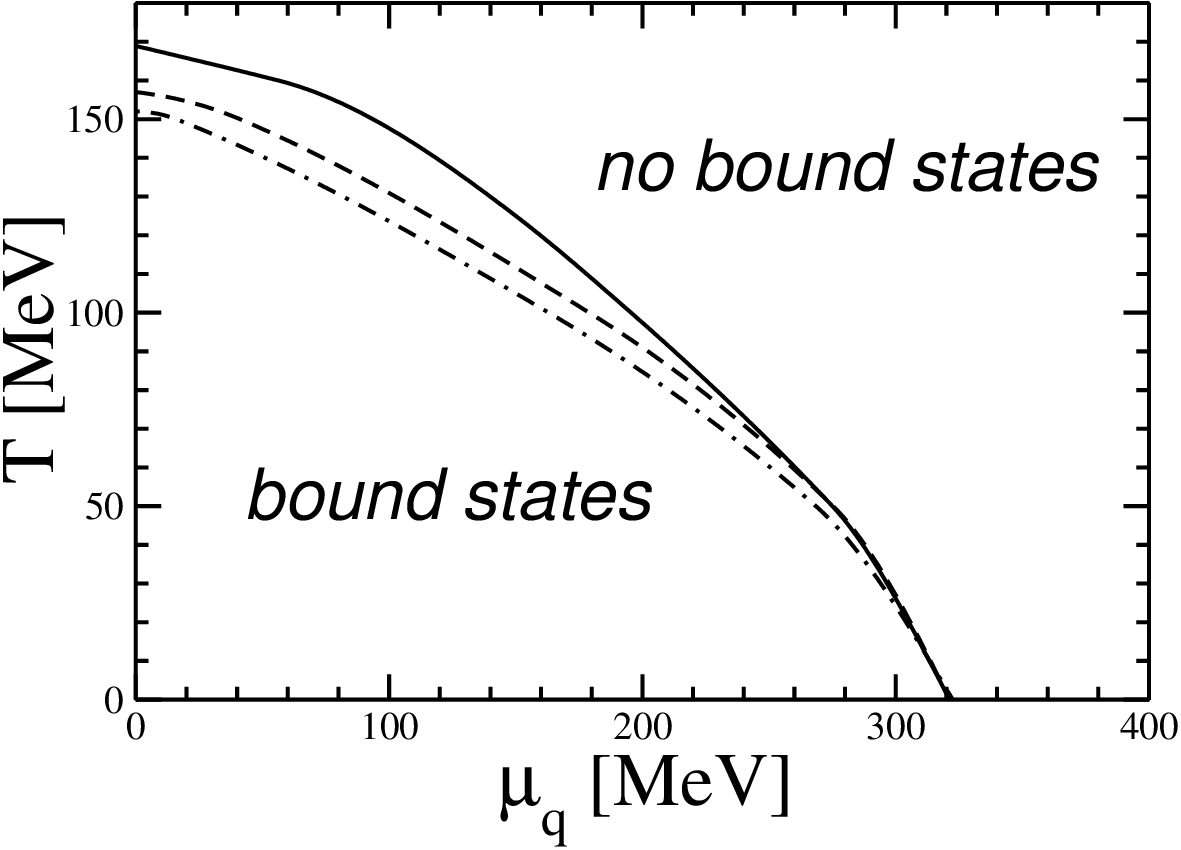}
\caption{Mott lines of the three-quark bound state as for different invariant
  cut-offs: $\Lambda=4m$ (solid),
  $\Lambda=6m$ (dash) $\Lambda=8m$ (dash-dot).\label{fig:phase}}
\end{minipage}
\hfill
\begin{minipage}[t]{0.48\textwidth}

\includegraphics[width=\textwidth]{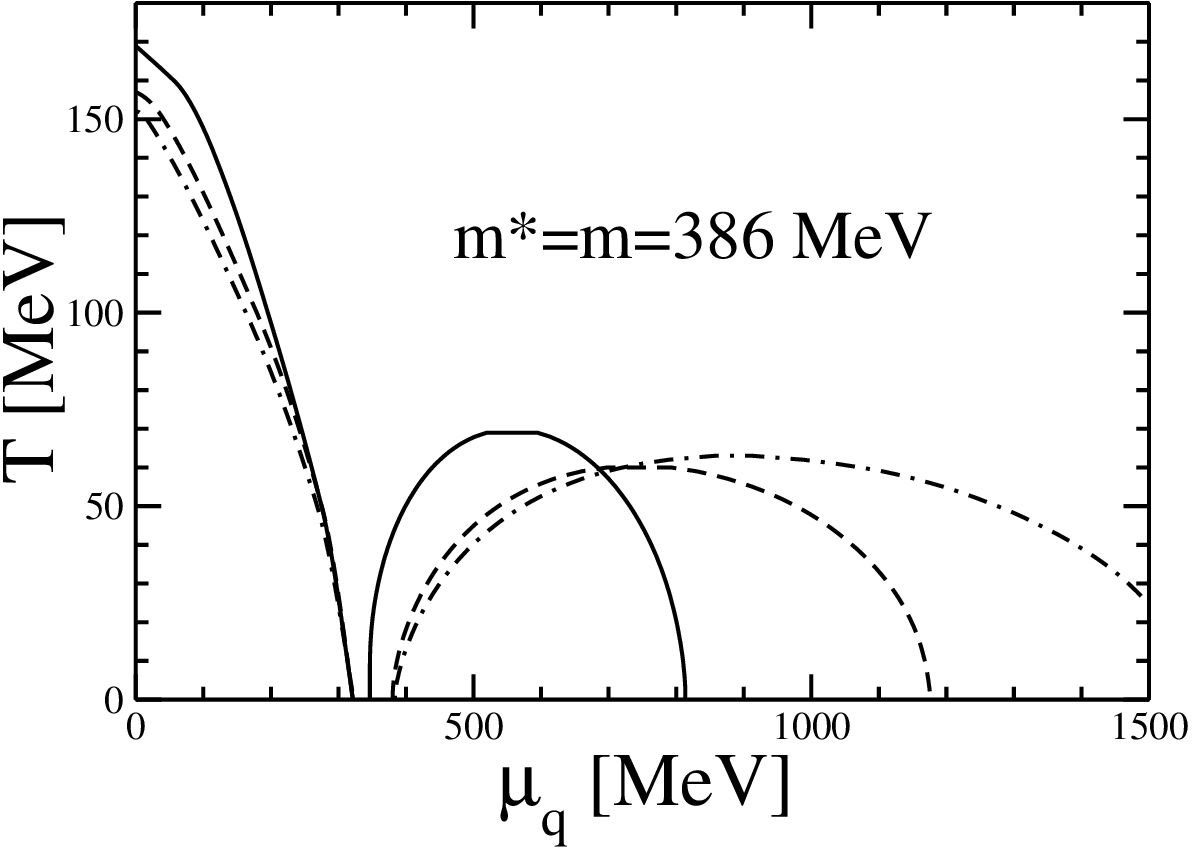}
\caption{Critical temperature of the color superconductivity for different invariant cut-offs: $\Lambda=4m$ (solid),
  $\Lambda=6m$ (dash) $\Lambda=8m$ (dash-dot).\label{fig:SC}}
\end{minipage}
\end{figure}

Using the Dyson approach we have derived consistent
relativistic few-body equations for particles embedded in a medium of
both finite temperature and finite density. They systematically include the effects of self energy corrections $m=m(T,\mu)$ and Pauli blocking factors, given in terms of the Fermi distribution functions. For particles they are~\cite{LFT}
\begin{eqnarray}
 f(k^+,\vec k^2_\perp)&=&\left(\exp\frac{1}{k_\mathrm{B}T}\left[\left(\frac{\vec k^2_\perp+m^2}{2k^+}+\frac{k^+}{2}-\mu\right)\right]+1\right)^{-1}
\label{eqn:Fermi}
\end{eqnarray}
expressed in terms of light front form momenta given by $\vec
k_\perp=(k_x,k_y)$ and $k^\pm=k_0\pm k_z$. The three-quark equation becomes
\begin{eqnarray}
\Gamma_\Lambda(y,\vec q_\perp) &= &\frac{i}{(2\pi)^3}\ t_\Lambda(M_2)
\int_0^{1-y} \frac{dx}{x(1-y-x)}\\
&&\hspace*{-1.5cm}\int d^2k_\perp
\frac{\theta(M^2_{30}-\Lambda^2)\left(1-f(x,\vec k^2_{\perp})-f(1-x-y,(\vec k+\vec q)_\perp)^2)\right)}
{M^2_3 -M_{03}^2}\;\Gamma_\Lambda(x,\vec k_\perp)\nonumber.
\label{eqn:med3}
\end{eqnarray}
Solving eq.(\ref{eqn:med3}) we calculate the three-quark binding energy as function of the temperature and the chemical potential for the different cut-offs.
We can go further and estimate at which temperature and chemical potential the binding energy goes to zero and therefore the three-quark bound states disappear (dissociation). The values of $T$ and $\mu$ for which this transition occurs are called Mott lines and are shown in Fig.~\ref{fig:phase}.
Their behavior qualitatively reflects the hadronic-deconfined phase transition.
At low temperatures the dependence on the cut-off is mild, but we note that at zero density the different $\Lambda$ give different values for the critical temperature. We can choose the cut-off by fitting our results to the lattice calculations.

Utilizing the Thouless criterion~\cite{Thoul} we calculate values of the critical temperature  $T_{\mathrm{c}}$ of the color superconducting-phase for the different cut-offs, using a medium-independent quark mass.
The preliminary results, shown in in Fig.~\ref{fig:SC}, present the qualitative behavior 
known from different model simulations.
The question if the onset of the color superconductivity at $T=0$ occurs in the deconfined phase is still open and further investigations of the chiral restoration may answer it.

\section{Conclusion} 
In conclusion, we have presented relativistic equations of the three-body problem using a zero range interaction and investigated its stability in isolated and in-medium cases.
Utilizing a invariant cut-off $\Lambda$ we have shown the relativistic analog of the Thomas collapse ($M_{3B}\rightarrow 0$). We find that the proton can be described as a Borromean state in case of weakly bound systems.
We have derived consistent relativistic three-quark equations at finite density and temperature. We find that the and the critical temperature for the color superconductivity agree qualitatively with results expected from other sources. Further analysis including a treatment of
spins and the chiral restoration are left for future investigations.

\begin{acknowledge}
I am grateful to M. Beyer, T. Frederico and H.J. Weber for the a fruitful collaboration.
Work supported by Deutsche Forschungsgemeinschaft.
\end{acknowledge}


\begin{thebibliography}{99}
\bibitem{Fedorov:2001wj}
Fedorov D.V. and Jensen A.S.:
Nucl.\ Phys.\ A {\bf 697}, 783 (2002)
\bibitem{Fedorov:2001vw}
Fedorov D.V. and Jensen A.S.: Phys.\ Rev.\ A {\bf 63}, 063608 (2001)
\bibitem{Frederico:1992uw}
Frederico T.: Phys.\ Lett.\ B {\bf 282}, 409 (1992)
\bibitem{Carbonell:2002qs}
Carbonell J. and Karmanov V.A:
Phys.\ Rev.\ D {\bf 67}, 052001 (2003)

\bibitem{Thomas1935}
Thomas L.H.: Phys. Rev. {\bf 47},903 (1935)
\bibitem{deAraujo:1995mh}
de Araujo W.R., de Melo J.P. and Frederico T.: Phys.\ Rev.\ C {\bf 52}, 2733 (1995)
\bibitem{Karsch:2000vy}
Karsch F.:
Nucl.\ Phys.\ Proc.\ Suppl.\ {\bf 83}, 14 (2000)
\bibitem{Alford:2001dt}
Alford M.:
hep-ph/0102047.
\bibitem{Beyer:2001bc}
Beyer M., Mattiello S., Frederico T. and Weber H.J.:
Phys.\ Lett.\ B {\bf 521}, 33 (2001)
\bibitem{Mattiello:2001vq}
Mattiello S., Beyer M., Frederico T. and Weber H.J.:
Few Body Syst.\  {\bf 31}, 159 (2002)
\bibitem{duk98}
Dukelsky J., R\"opke G., and Schuck P.: Nucl.\ Phys.\  {\bf A628}, 17 (1998)
\bibitem{stab:2003}
Beyer M., Mattiello S., Frederico T. and Weber H.J.:
Few Body Syst.\  {\bf 33}, 89 (2003)
\bibitem{LFT}
Beyer M., Mattiello S., Frederico T. and Weber H.J.:
arXiv:hep-ph/0310222
\bibitem{Thoul}
Thouless D.J.: Annals Phys.\ {\bf 10}, 553 (1960)
\end{thebibliography}
\end{document}